\pdfoutput=1
\documentclass[aps,pra,reprint,amsmath,amssymb,superscriptaddress]{revtex4-1}
\usepackage{bm}
\usepackage{graphicx}
\usepackage[colorlinks]{hyperref}
\usepackage{mathrsfs}
\usepackage{times}

\DeclareMathOperator\erf{erf}

\newcommand{\beq}{\begin{equation}}
\newcommand{\eeq}{\end{equation}}

\begin{document}

\title{Thermodynamics of two-dimensional bosons in the lowest Landau level}

\author{Bhilahari Jeevanesan}
\affiliation{Department of Physics, Technical University of Munich, 85748 Garching, Germany}
\affiliation{Munich Center for Quantum Science and Technology (MCQST), Schellingstr. 4, D-80799 M\"unchen}
\author{Sergej Moroz}
\affiliation{Department of Physics, Technical University of Munich, 85748 Garching, Germany}
\affiliation{Munich Center for Quantum Science and Technology (MCQST), Schellingstr. 4, D-80799 M\"unchen}

\begin{abstract}% maximum 600 characters!
We study the thermodynamics of short-range interacting, two-dimensional bosons constrained to the lowest Landau level. When the temperature is higher than other energy scales of the problem, the partition function reduces to a multidimensional complex integral that can be handled by classical Monte Carlo techniques. This approach takes the quantization of the lowest Landau level orbits fully into account. We observe that the partition function can be expressed in terms of a function of a single combination of thermodynamic variables, which allows us to derive exact thermodynamic relations. We determine the asymptotic behavior of this function and compute some thermodynamic observables numerically.
\end{abstract}

\maketitle

%%%%%%%%%%%%%%%%%%%%%%%%%%%%%%%%%
%%%%%%%%%%%%%%%%%%%%%%%%%%%%%%%%%

\section{Introduction}
\label{sec:intro}
Two-dimensional quantum matter often responds to an external magnetic flux with the generation of point-like vortices. Apart from the celebrated Abrikosov vortices in type-II superconductors, realizations of such states can also be found in neutral, harmonically trapped superfluids, where one can artificially mimic a magnetic field by rotating the fluid \cite{Fetter2009, svistunov2015superfluid,sonin2016dynamics}. With interactions included, the fate of such a system at zero temperature is determined by its filling fraction $\nu$ \cite{Cooper2008, Sonin2016}, defined as the ratio of the density of bosons $n$ to the density of vortices $n_v=B/2\pi$. At large fillings, which for short-range isotropically interacting bosons is $\nu \gtrsim 8$ \cite{Cooper2001, Sinova2002}, a compressible superfluid phase is formed with vortices arranged into a periodic crystal. At small fillings the vortex crystal melts and at some special fillings bosons form incompressible, strongly-correlated fractional quantum Hall states \cite{Cooper2008, Viefers2008}. 

At sufficiently large temperatures a vortex crystal undergoes a thermal melting transition. 
The thermodynamics of vortex matter and the nature of the melting transition was mainly discussed in the context of type-II superconductors within the bosonic Ginzburg-Landau phenomenological theory. The mean-field theory of Abrikosov predicts a continuous second-order phase transition from the vortex crystal to a normal state  \cite{Abrikosov1957}. It was demonstrated in \cite{Brezin1985}, however, that fluctuations should invalidate this picture and render the transition first order, at least close to the upper critical dimension $d=6$. On the other hand, in thin superconducting films it was proposed \cite{Fisher1980} that the vortex crystal melts into a vortex fluid via a pair of Berezinskii-Kosterlitz-Thouless (BKT) like phase transitions  arising from the unbinding of crystal dislocations and disclinations. 
In the past, a number of studies of the thermodynamics of the classical Ginzburg-Landau model have been undertaken \cite{Hikami1991, *Tesanovic1992, *Kato1993, *ONeil1993} leading to contradictory results for the order of the vortex crystal melting transition. Recent experimental work with very weakly pinned superconducting films supports the two-stage BKT melting scenario\cite{Roy2019}.

In rotating two-dimensional bosonic superfluids 	the defect-mediated BKT melting of the vortex crystal was discussed in \cite{Gifford2008}.
Numerical studies of the thermodynamics of quantum bosons in magnetic fields are hampered by the fact that quantum Monte Carlo simulations suffer from the notorious sign problem.

\begin{figure}[h] 
\centering{}\includegraphics[width= 7.0 cm]{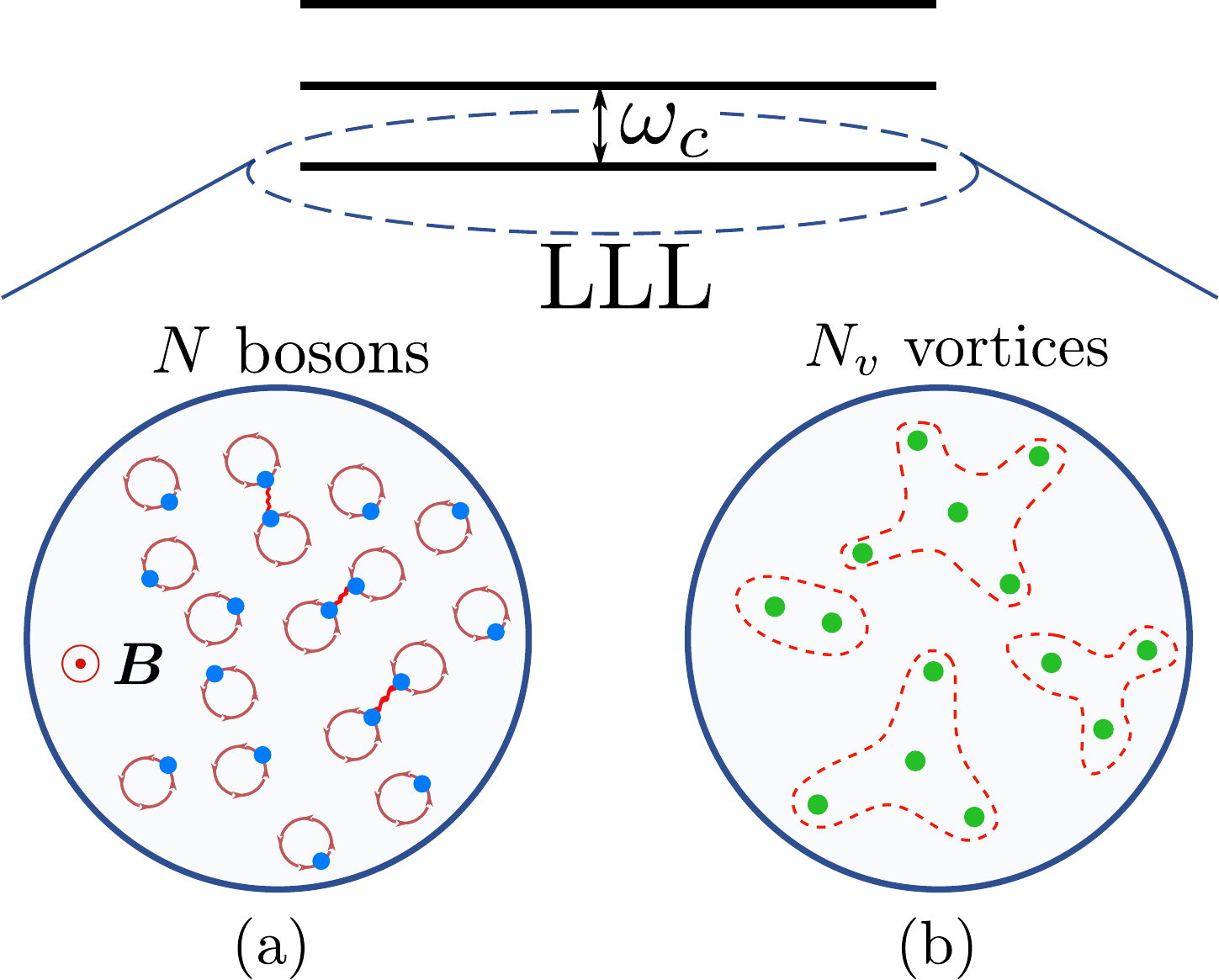}\protect\caption{\label{fig:firstFig} The physics of a bosonic droplet in the lowest Landau level (LLL). (a) The magnetic field $B$ constrains $N$ bosons (blue) to occupy cyclotron orbits. The bosons interact by a two-body short-range interaction and repel each other when close. (b) Alternatively, one may view the system as one of $N_v$ interacting vortices (green). The interaction between vortices is not pairwise, but a sum of multi-body interactions \cite{PhysRevA.76.053602}, which are indicated by broken lines enclosing multiple vortices. At low temperatures the vortices freeze into an Abrikosov lattice of size $\sim \sqrt{N_v}l_B$ in order to minimize the interaction energy.}
\end{figure}

In this paper we investigate theoretically the thermodynamics of a finite-size two-dimensional droplet of bosons placed into a magnetic field and restricted to the lowest Landau level (LLL), see Fig. \ref{fig:firstFig}.
In this setting the problem reduces to the quantum statistical physics of $N_v+1$ complex degrees of freedom, where $N_v$ is the number of quantum vortices. While previously this setup has been studied at zero temperature (for a review of this vast field, see \cite{Viefers2008, saarikoski2010vortices, Cooper2001} and cited literature therein), we focus our attention here on the regime where the temperature is higher than the remaining energy scales of the problem. In this limit we can neglect the quantum fluctuations of the vortices and derive a simpler model that we study in the remainder of this paper. We observe that the partition function of this model is fully specified by a function $\psi$ of only one variable, see eq. \eqref{eq:RescaledpartitionFct}. We use this constrained form to derive exact relations \eqref{eq:exactFormulaforEN} and \eqref{eq:exactFormulaforMagnetization} between thermodynamic observables. With the help of classical Monte Carlo simulations we map out the function $\psi$ and determine its asymptotics \eqref{eq:psiAsymptotics}. We make use of the latter to obtain the asymptotics of thermodynamic observables.

Recent advances in cold atom experiments allowed the thermodynamics of bosonic and fermionic superfluids to be determined with unprecedented accuracy \cite{Nascimbene2010, *Yefsah2011, *Ku2012, *Desbuquois2014}. Our work provides motivation to measure the equation of state and other thermodynamic observables of two-dimensional bosonic superfluids in effective magnetic fields. The latter have been realized by external rotation or with the use of artificial gauge fields, for reviews see \cite{Bloch2008} and \cite{Aidelsburger2018}.
While ultracold atom experiments exploring rotating Bose gases were already done decades ago \cite{schweikhard2004rapidly, Abo2001}, the limit of the LLL is reached explicitly in very recent works \cite{Fletcher2019, Chalopin2020} using new experimental ideas.
%%%%%%%%%%%%%%%%%%%%%%%%%%%%%%%%%

\section{Thermodynamics in the lowest Landau level}
\label{sec:LLLThermo}

The system that we study is a two-dimensional droplet of identical bosons in a constant magnetic field $B$ that points perpendicular to the plane. The bosons interact with each other by means of a contact repulsive potential. We are interested in the thermodynamics of this system and to this end consider the grand canonical quantum partition function expressed in terms of a functional field integral
$ Z  = \int D[\bar{\phi},\phi] e^{-S[\bar{\phi},\phi] }$
with $S=\int\limits_{0}^{\beta} d\tau \int d^2 x \mathcal{L}$ and
\beq
\label{eq:action_1}
\mathcal{L}=  \   \bar\phi \partial_\tau \phi +  \frac{1}{2m} \bar \phi(-i\nabla - {\bf A})^2 \phi - \mu \bar \phi \phi +\frac g 2 \left( \bar \phi \phi \right)^2. 
\eeq
Here $\beta$ is the inverse of the temperature $T$, $m$ is the mass of the bosons, $\mu$ is the chemical potential which we tune to fix the number of bosons in the system and $\bm A$ is the vector potential corresponding to the magnetic field $ B$. We choose to work in the symmetric gauge, for which $\bm A = (- By/2, Bx/2)$. In this paper we set $\hbar= k_B = 1$ and absorb the electric charge of the bosons into the magnetic field.

Since bosons are subject to a constant magnetic field, one can expand the complex field $\phi$ in terms of Landau level eigenfunctions. The spacing between consecutive levels is given by the cyclotron frequency $\omega_c = B/m$. When the interaction energy per particle $g n$ and the temperature $T$ are much smaller than $\omega_c$, we can restrict our attention only to the LLL. Formally this can be realized by taking the limit $\omega_c \rightarrow \infty$ and $m \rightarrow 0$ such that $B\sim n_v$  remains finite. 
In this limit the contributions of the higher Landau levels to the action vanish and we have effectively restricted the functional integral to states in the LLL. 
Thus the bosonic field can be expanded in terms of normalized LLL eigenfunctions
\begin{eqnarray} \label{LLLphi}
\phi(z, \tau) &=& \sum_{n = 0}^{N_v} \frac{c_n(\tau)}{\sqrt{2^{n+1}\pi n!}l_B^{n+1}} z^n e^{-|z|^2 /4l_B^2} \nonumber\\
&\equiv& C(\tau) \prod_{i = 1}^{N_v}[z - z_i(\tau)] e^{-|z|^2 /4l_B^2}
\end{eqnarray}
where $z = x + i y$ and the $c_n(\tau)$ are dimensionless complex coefficients that depend on imaginary time and satisfy $c_n(\beta)=c_n(0)$. The quantity $l_B$ is the magnetic length and equals ${1}/{\sqrt B}$. The degeneracy of the LLL is $N_v +1$ and it is related to the area $A$ of the droplet as $N_v \sim A / l_B^2 $. Clearly the LLL function $\phi(z,\tau)$ has $N_v$ zeros, which are the positions of the vortices, denoted by $z_i(\tau)$ in equation \eqref{LLLphi}. \newline
Inserting the form \eqref{LLLphi} into the action and carrying out the spatial integration, the partition function takes the form $Z = \int D[\{\bar{c}(\tau),c(\tau)\}] e^{-S[\{\bar{c},c\}]}$ with the Lagrangian being
\begin{eqnarray} \label{LLLQuantumaction}
\mathcal{L}&=&  \   \sum_{n = 0}^{N_v}\bar c_n \left(\partial_\tau - \mu\right) c_n \nonumber \\
&&+ \frac{g}{4 \pi l_B^2} \sum_{s = 0}^{2 N_v} \left| \sum_m \sqrt{2^{-s}{s \choose m}} c_m(\tau) c_{s-m}(\tau)\right|^2, 
\end{eqnarray}
where in the last line the sum runs over all $m$ with $0\leq m \leq N_v$ and $0\leq s - m \leq N_v$. The chemical potential has been shifted to absorb the constant energy $\omega_c/2$ resulting from the gradient term \footnote{At the cost of a Vandermonde determinant one could describe the partition function in terms of vortex coordinates instead of the coefficients $\{c_n\}$. However, in this case the contact interaction between bosons leads to a rather complicated multivortex-interaction \cite{PhysRevA.76.053602}. For this reason we prefer to work within the $\{c_n\}$-description, and only make reference to vortices when discussing the vortex crystal at low temperatures.}. \newline
The Lagrangian in \eqref{LLLQuantumaction} contains the two energy scales $|\mu|$ and $g' = g/(4\pi l_B^2)$. If the temperature is much higher than both of these scales, i.e. $T \gg |\mu|,  g'$, we are justified in neglecting quantum fluctuations of the $\{c_n\}$ and treating the problem classically. Thus in this high-temperature limit we make the static approximation $c_n(\tau)\to c_n$ giving rise to the partition function of the form 
\begin{eqnarray}
\label{eq:finalpartitionFct}
 Z  =  \prod_{n = 0}^{N_v}{\int\frac{d\bar{c_n}dc_n}{2\pi}} e^{- \beta H[\{\bar{c},c\}] }
\end{eqnarray}
with 
\begin{eqnarray}
\label{eq:finalHamiltonian}
H[\{\bar{c},c\}] &=&  - {\mu} \sum_{n = 0}^{N_v} \bar{c}_n c_n \nonumber \\
& & + \frac{g}{4 \pi l_B^2} \sum_{s = 0}^{2 N_v} \left| \sum_m \sqrt{2^{-s}{s \choose m}} c_m c_{s-m}\right|^2.
\end{eqnarray}
We emphasize that despite the seemingly classical form of the partition function \eqref{eq:finalpartitionFct}, quantum mechanics enters in this approach due to the restriction to the lowest Landau level. Formally the Hamiltonian \eqref{eq:finalHamiltonian} depends on the magnetic length $l_B$, which is the quantum mechanical length scale fixing the size of cyclotron orbits. 
%%%%%%%%%%%%%%%%%%%%%%%%%%%%%%%%%

\subsection{Exact relations}
\label{sec:Exact}
We find that, up to a known factor, the classical partition function \eqref{eq:finalpartitionFct} is a function of a single dimensionless variable formed out of the temperature $T$, chemical potential $\mu$ and the modified interaction strength $g' = g/(4\pi l_B^2)$. To show this we rescale in eq. \eqref{eq:finalpartitionFct} all coefficients by $c_n \rightarrow \sqrt[\leftroot{-2}\uproot{2}4]{T/g'} c_n$. Apart from changing the terms in the Hamiltonian (\ref{eq:finalHamiltonian}), this rescaling also affects the integration measure in (\ref{eq:finalpartitionFct}). This brings the partition function into the form
\begin{eqnarray}
\label{eq:RescaledpartitionFct}
Z  =  \left(\frac{T}{g'}\right)^{\frac{N_v +1}{2}}\psi\left(x\right),
\end{eqnarray}
where $x=\mu/\sqrt{g' T}$.\newline
First we explore general consequences which this functional form entails. The average particle number is obtained from the thermodynamic potential $\Omega = -T \log  Z$  by $\langle N \rangle = - \partial_{\mu} {\Omega}$, i.e.
\begin{eqnarray}
\label{eq:FormulaforN}
\langle N \rangle  =  \sqrt{\frac{T}{g'}} \frac{\psi'\left(x\right)}{\psi\left(x\right)},
\end{eqnarray}
where the prime on $\psi$ denotes differentiation with respect to $x$.
Meanwhile, the average energy is $\langle E \rangle = -\partial_\beta \log  Z + \mu \langle N \rangle$ and thus
\begin{eqnarray}
\label{eq:FormulaforE}
\langle E \rangle = \frac{N_v + 1}{2} T + \frac{\mu}{2} \sqrt{\frac {T} {g'}} \frac{\psi'\left(x\right)}{\psi\left(x\right)}.
\end{eqnarray}
Eliminating the ratio $\psi'/\psi$ from both equations, we obtain an exact relation between $\langle N \rangle$ and $\langle E \rangle$
\begin{eqnarray}
\label{eq:exactFormulaforEN}
\langle E \rangle = \frac{N_v + 1}{2} T  + \frac{\mu }{2} \langle N \rangle.
\end{eqnarray}
Along similar lines one can derive a universal relation that relates the magnetization $ \langle M \rangle=-\partial_B \Omega$ to the particle number $\langle N \rangle$. The magnetic field enters the partition function only through the magnetic length $l_B = 1/\sqrt{B}$, thus $\partial_B = g/(4\pi) \partial_{g'}$. After carrying out the derivatives one finds the relation 
\begin{eqnarray}
\label{eq:exactFormulaforMagnetization}
\langle E \rangle = -B \langle M \rangle.
\end{eqnarray}
%%%%%%%%%%%%%%%%%%%%%%%%%%%%%%%%%

\subsection{Determination of $\psi(x)$ from Monte Carlo simulations}
The fact that the partition function has a simple functional form means that one can extract the function $\psi(x)$ by following a specific curve in  the $T$-$\mu$ parameter space. In the following we present results obtained from Monte Carlo simulations, where we calculate the average particle number $\langle N \rangle$ at fixed $T = g'$ and for varying values of $\mu$. As seen from eq. (\ref{eq:FormulaforN}), on this particular trajectory in $T$-${\mu}$ space $\langle N(x) \rangle$ reduces to $\langle N(x) \rangle = \psi'(x)/\psi(x)$ and by calculating the left-hand-side numerically, we can obtain the function $\psi(x)$ up to an overall multiplicative constant. We now briefly describe the Monte Carlo simulation. Starting from randomized initial values for the coefficients $c_n$, we repeatedly update their values in the complex plane. A proposal for an update is either accepted or rejected according to the standard Metropolis rule. The size of the change of $c_n$ is chosen uniformly randomly inside a  circle of a certain radius. This radius is chosen such that the ratio of accepted updates to proposed updates is around $0.4$. Then we repeat a cycle of thermalization and measurement, which typically have $10^5$ steps each. The number of cycles is $100$. 

\begin{figure}[h]
\centering{}\includegraphics[width=\columnwidth]{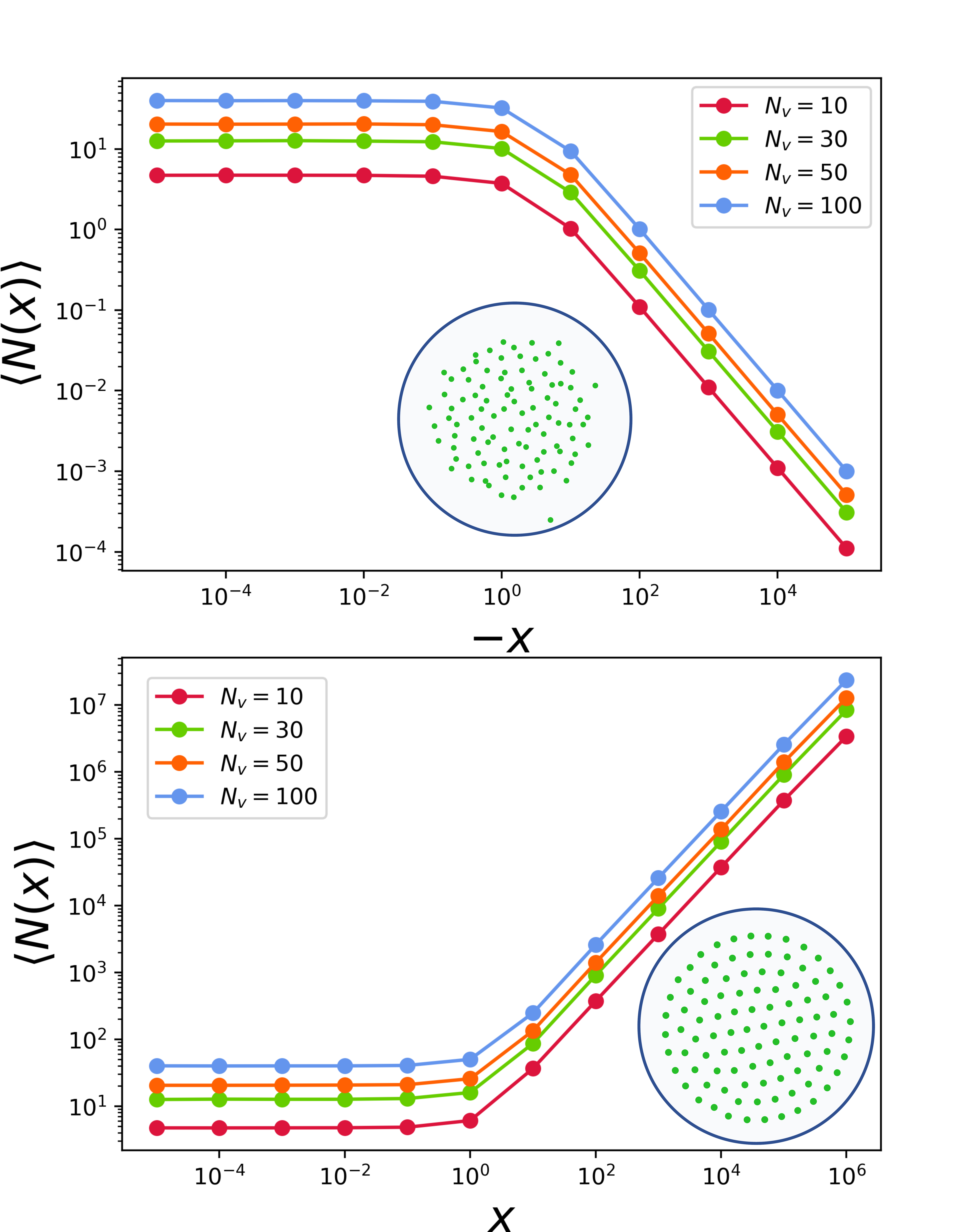}\protect\caption{\label{fig:Npowerlaws} Monte Carlo result for the particle number $\langle N \rangle$ at fixed temperature $T = g'$ and chemical potential $\mu= x g'$ varying with $x$. Each plot also shows a snapshot of the vortex positions that is characteristic for the value of $x$ taken from the simulation. The vortex positions $z_i$ are found from the coefficients $c_n$ by factorizing the polynomial part of $\phi$ as shown in eq. (\ref{LLLphi}). The region of large negative values of $x$ (top) corresponds to the high temperature limit, where the vortices form a fluid, whereas values of large positive $x$ (bottom) correspond to low temperatures, at which one finds the vortices to be arranged into a crystal. A video showing our simulation is found in \cite{supmat}.}
\end{figure}

The two plots in Fig. \ref{fig:Npowerlaws} show $\langle N(x) \rangle$ in doubly-logarithmic form for negative and positive values of $x$. At large and small $|x|$ the observable $\langle N(x) \rangle$ clearly follows power laws and we determine the exponents by fitting. We find 

\begin{eqnarray}
\label{eq:NAsymptotics}
\langle N(x) \rangle =\begin{cases}
-\frac{c}{x} & x  \rightarrow -\infty\\
b & x \rightarrow 0 \\
a x & x \rightarrow +\infty .  \end{cases}
\end{eqnarray}

The parameters $a, b$ and $c$ are functions of $N_v$. For large $N_v$ they are well described by $a \approx 0.24 (N_v + 1)$, $b \approx 0.4 (N_v + 1) $ and $c = N_v +1$.
We can now obtain asymptotic formulas for $\psi(x)$ by solving the differential equation (\ref{eq:FormulaforN}) with (\ref{eq:NAsymptotics}) as input. We obtain 

\begin{eqnarray}
\label{eq:psiAsymptotics}
\psi(x) \sim \begin{cases}
\frac{1}{\left|x\right|^{N_v +1}} & x  \rightarrow -\infty \\
e^{{b x}} & x \rightarrow 0 \\
e^{{\frac{a}{2} x^2}-p \log x}  & x \rightarrow +\infty,  \end{cases}
\end{eqnarray}
up to multiplicative constants. The limit $x\rightarrow -\infty$ corresponds to the non-interacting case, $g'\rightarrow 0$, and was therefore computed exactly. In the third case we have anticipated a subleading factor $x^{-p}$ that we will discuss in the context of the specific heat below. 
%%%%%%%%%%%%%%%%%%%%%%%%%%%%%%%%%

\subsection{Thermodynamic observables}
\begin{figure}[h]
\centering{}\includegraphics[width=\columnwidth]{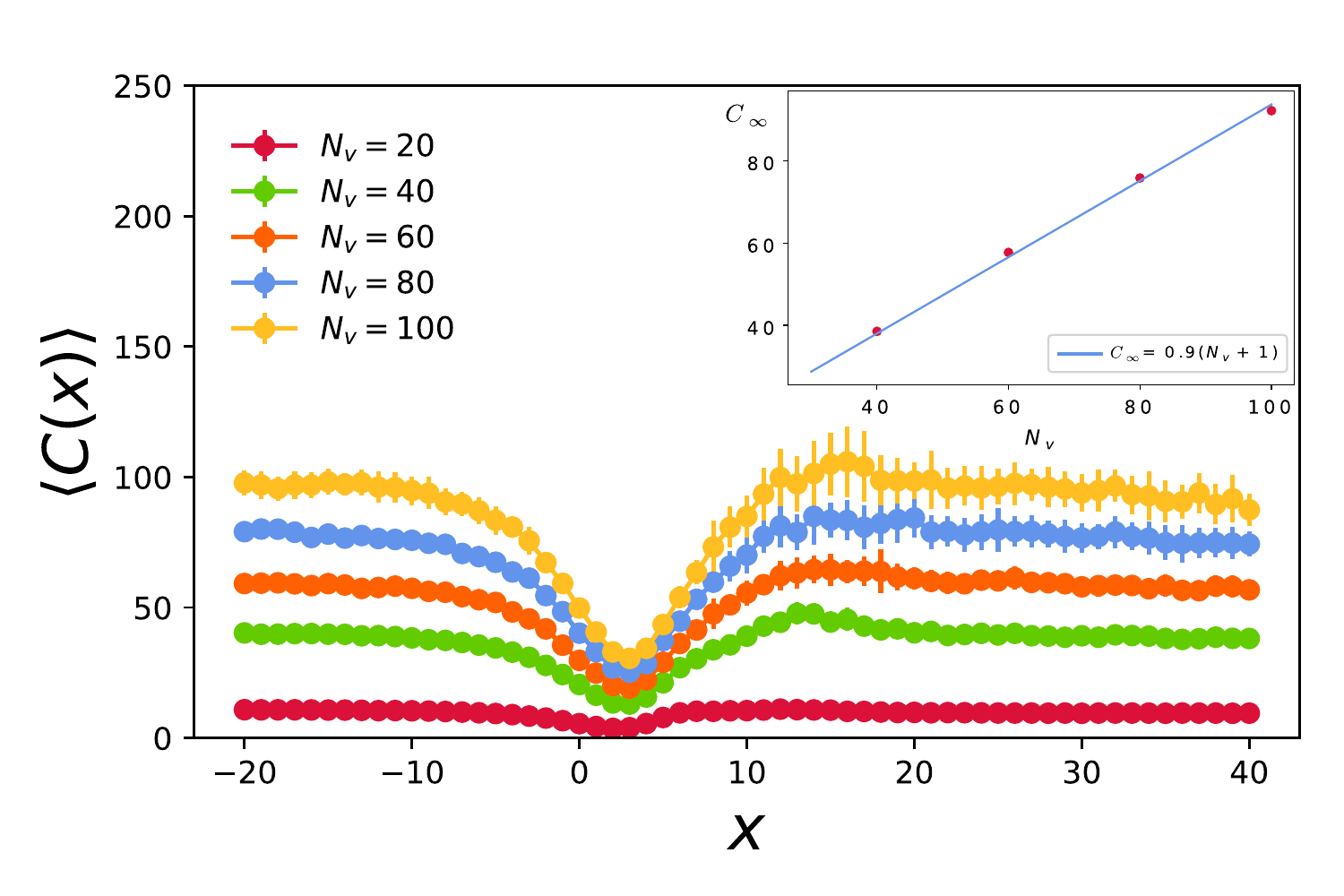}\protect\caption{\label{fig:C_curves} Monte Carlo result for the specific heat $C=-T \partial^2_T \Omega $ as a function of $x$. The curves reach a plateau at large $x$. The plateau height is shown in the inset as a function of $N_v \gg 1$.}
\end{figure}

The Monte-Carlo simulation was done along the trajectory $(g',xg')$ in $T$-$\mu$ space by changing $x$. With the expressions for $\psi(x)$ in hand, we can now obtain further knowledge about $\langle N \rangle$ for general $T$ and ${\mu}$:
\begin{eqnarray}
\label{eq:FullNAsymptotics}
\langle N(T,{\mu},g') \rangle =\begin{cases}
-{(N_v +1)T}/{{\mu}}  & g' \rightarrow 0, \mu < 0\\
b \sqrt{{T}/{g'}} & {\mu} \rightarrow 0 \\
a{{\mu}}/{g'} - p T/\mu & T \rightarrow 0, \mu > 0.  \end{cases}
\end{eqnarray}
%Similarly, the expression for the energy of the system in terms of $\psi(x)$ in eq. (\ref{eq:FormulaforE}) yields:	
%\begin{eqnarray}
%\label{eq:FullEAsymptotics}
%\langle E(T,{\mu},g') \rangle =\begin{cases}
%0 & g'  \rightarrow 0, \mu < 0\\
%\frac{N_v+1}{2}T+\frac{b}{2}\sqrt{\frac{T}{g'}} {\mu} & {\mu} \rightarrow 0 \\
%\frac{a\mu^2}{2g'} + \frac{N_v +1 - p}{2}T & T \rightarrow 0, \mu > 0.  \end{cases}
%\end{eqnarray}
\newline The specific heat at constant chemical potential $C = T\partial_T S =-T \partial^2_T \Omega $ is obtained from the Monte Carlo simulation by calculating the correlator $C=\beta^2\left[\langle \left(E-\mu N\right)^2 \rangle - \langle E-\mu N \rangle^2 \right]$.
In terms of the function $\psi(x)$, the specific heat can also be expressed as
\begin{eqnarray}
\label{eq:SpecificHeatFormula}
C = \frac{N_v + 1}{2} - \frac{x}{4} \frac{\psi'(x)}{\psi(x)} + \frac{x^2}{4} \left(\frac{\psi'(x)}{\psi(x)}\right)'.
\end{eqnarray}
Clearly this quantity does not depend on the specific trajectory chosen, since it depends only on $x$. \newline
The result of the Monte Carlo simulation is shown in Fig. \ref{fig:C_curves}.

\begin{figure}[h]
\centering{}\includegraphics[width=\columnwidth]{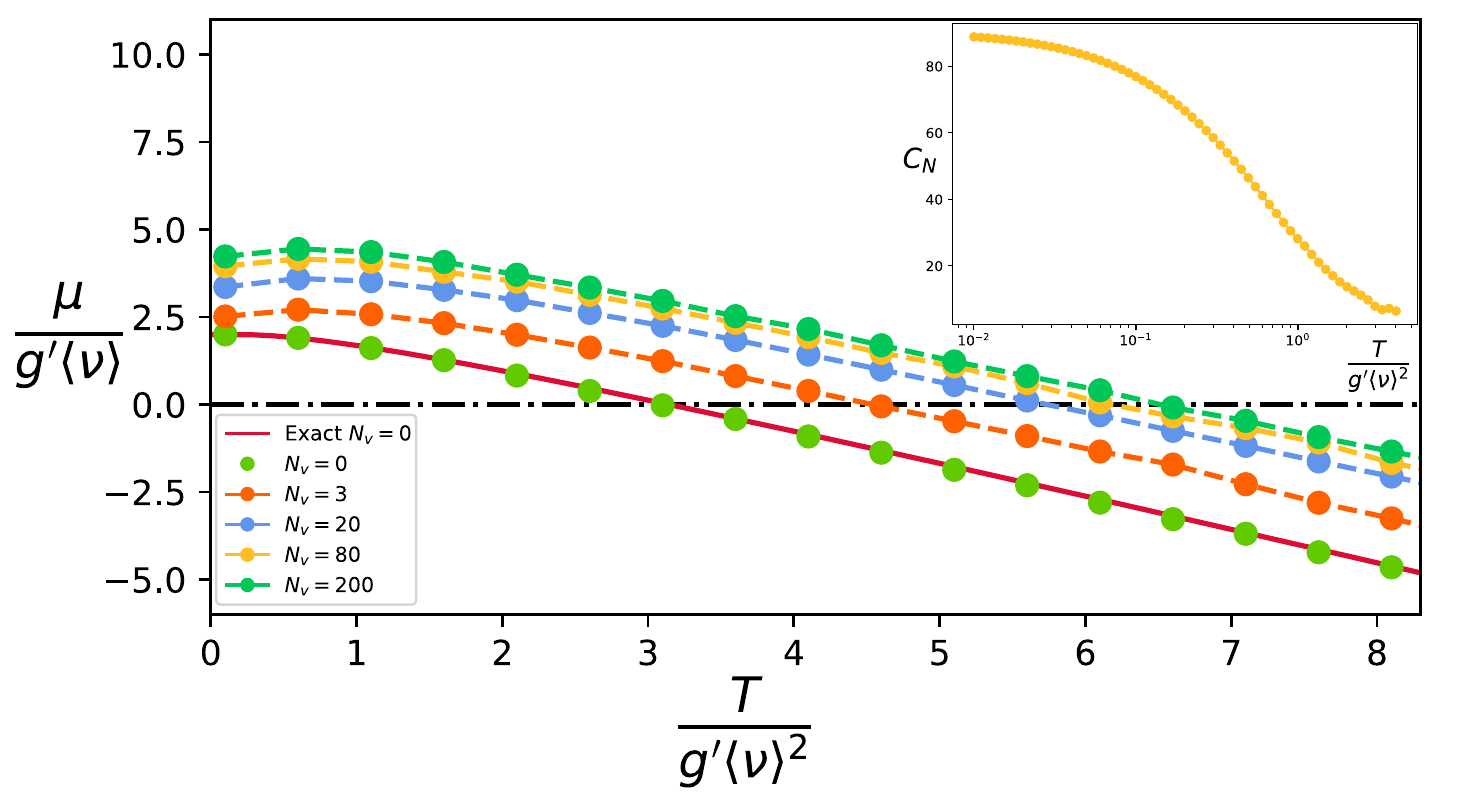}\protect\caption{\label{fig:muvsT} The chemical potential $\mu$ as a function of temperature at fixed particle number $\langle N \rangle = 10^4$. Here $\langle \nu\rangle  = \langle N\rangle /(N_v +1)$ is the filling fraction defined in the text. The curve for zero vortices is exact, see appendix~\ref{app:Exact}, while the other lines were obtained from the Monte Carlo simulation. Using this data we can find the specific heat at constant particle number, see appendix~\ref{app:SpecHeat}. The inset shows such a plot for $N_v = 80$.}
\end{figure}

Inserting the asymptotics \eqref{eq:psiAsymptotics} in \eqref{eq:SpecificHeatFormula} we find predictions for the specific heat:
\begin{eqnarray}
\label{eq:CAsymptotics}
\langle C \rangle =\begin{cases}
N_v + 1 & x  \rightarrow -\infty\\
\frac{N_v + 1}{2} - \frac{b}{4} x & x \rightarrow 0 \\
\frac{N_v + 1 + p}{2} & x \rightarrow +\infty   \end{cases}
\end{eqnarray} 
At large $x$ the specific heat approaches a plateau $C_\infty = (N_v + 1 +p)/2$. As follows from the inset of Fig. \ref{fig:C_curves}, for large $N_v$ one has $p \approx 0.8(N_v +1)$. The region of large positive $x$ corresponds to low temperatures, as a consequence the specific heat of the vortex crystal tends to a constant in our model when $T\rightarrow 0$. This is a well-known deficiency of the classical treatment \cite{FermiThermo} that stems from neglecting higher Matsubara components of the LLL degrees of freedom $\{c_n\}$. \newline 
The fact that $C(0) < C(\pm\infty)$, necessitates the existence of a global minimum of $C(x)$, which is the dip seen in Fig. $\ref{fig:C_curves}$.
\newline
The behavior of the chemical potential as a function of temperature at fixed particle number is shown in Fig. \ref{fig:muvsT}.
The main features of this figure can be understood with the help of eq. \eqref{eq:FullNAsymptotics}. For high temperatures the system behaves essentially as if it were non-interacting, thus according to the asymptotics for $\langle N\rangle$ the slopes of the curves in the figure tend to $-1$. The asymptotics also predicts where the curves cross the $\mu = 0$ axis, namely at ${T_\star = [(N_v + 1)/b]^2  g'\langle\nu\rangle^2}$. For large $N_v$ this becomes $T_\star \approx 6.3  g'\nu^2$, which is indeed where the $N_v = 100$ curve becomes zero in Fig. \ref{fig:muvsT}. Finally, the low temperature asymptotics for $\mu$ is ${\mu = g'\langle N \rangle /a + p T/{\langle N \rangle}}$. For large $N_v$ this yields the line $\mu/(g'\langle \nu \rangle) \approx 4 + 0.8 T/(g'\langle \nu \rangle^2) $, correctly reproducing the intercept and slope at $T = 0$ of the $N_v = 100$ curve.

With the aid of the $\mu(T)$ curves in  Fig. \ref{fig:muvsT} it is possible to derive the specific heat $C_N$ at constant particle number, see appendix \ref{app:SpecHeat} for details. The result is shown in the inset of  Fig.~\ref{fig:muvsT}. As a consequence of working in the LLL this specific heat vanishes at high temperatures.

\section{Outlook}
An exciting extension of this work would be to take quantum fluctuations of the LLL degrees of freedom fully into account, since this can shed some light on the low temperature properties of the vortex crystal and may even illuminate the question of the {quantum} melting transition \cite{Sinova2002, Cooper2008}. In the regime $N \gg N_v$ the approach of using the small number $N_v+1$ of LLL basis states compared to the large number $N$ of bosons implies that a quantum Monte Carlo simulation of the LLL Lagrangian \eqref{LLLQuantumaction} is actually much more feasible than the ab-initio simulation of interacting bosons in a magnetic field. 

In order to clarify the nature of the {thermal} melting transition from a vortex crystal to a vortex fluid it may be fruitful to study within our approach the hexatic order parameter. However, studies of melting transitions of two-dimensional solids revealed that an extremely large number of degrees of freedom become neccessary. As an example, the melting transition in the two-dimensional hard-disk model was only resolved with the invention of an algorithm that was fast enough to equilibrate $\sim 10^6$ hard disks \cite{PhysRevLett.107.155704}. We presume that a collective-move algorithm similar to the one used for hard disks may help uncover the nature of the thermal melting transition of the vortex crystal.

%

%%%%%%%%%%%%%%%%%%%%%%%%%%%%%%%%%

\begin{acknowledgements}
\emph{Acknowledgements.}---We acknowledge useful discussions with Nicolas Dupuis, Egor Kiselev, Dam Thanh Son and Wilhelm Zwerger.  Our work ~is funded by the Deutsche Forschungsgemeinschaft (DFG, German Research Foundation) under Emmy Noether Programme grant no.~MO 3013/1-1 and under Germany's Excellence Strategy - EXC-2111 - 390814868. 
\end{acknowledgements}
%%%%%%%%%%%%%%%%%%%%%%%%%%%%%%%%%

\appendix
%\setcounter{page}{1}
%\renewcommand{\theequation}{A\arabic{equation}}
%\renewcommand{\thefigure}{A\arabic{figure}}
%\setcounter{equation}{0}
%\setcounter{figure}{0}

%%%%%%%%%%%%%%%%%%%%%%%%%%%%%%%%%
\section{Specific heat at constant particle number}
\label{app:SpecHeat}

The entropy is obtained as usual from $S=\partial_T(T\log Z)$ and by inserting the specific form of the partition function \eqref{eq:RescaledpartitionFct}, we obtain the result
\begin{eqnarray}
\label{eq:genFormS}
S = \frac{N_v + 1}{2}\left(1+\log \frac{T}{g'}\right)  + \log \psi(x) - \frac{x}{2} \frac{\psi'(x)}{\psi(x)}.
\end{eqnarray}
The specific heat at constant chemical potential is found from this by $C=T\partial_T S \bigr \vert_\mu$, which yields
\begin{eqnarray}
\label{eq:genFormC}
C = \frac{N_v + 1}{2} - \frac{x}{4} \frac{\psi'(x)}{\psi(x)} + \frac{x^2}{4} \left(\frac{\psi'(x)}{\psi(x)}\right)'.
\end{eqnarray}
At high temperatures we find from the asymptotics \eqref{eq:psiAsymptotics} that the entropy behaves like
\begin{eqnarray}
S&=&\frac{N_v + 1}{2} + (N_v + 1) \log \left (-\frac{T}{\mu} \right) \\
&=& \frac{N_v + 1}{2} + (N_v + 1) \log \left (\frac{N}{N_v + 1} \right),
\end{eqnarray}
where we used the fact that at high temperatures ${\mu = -(N_v + 1)/N \times T}$. Thus the entropy of a system with $N$ particles tends to a constant as $T\rightarrow \infty$, despite the fact that the specific heat $C$ tends to a constant. 

In the main text the $\mu(T)$ curve was determined for fixed particle number. This function allows us to determine $C_N$, the specific heat at constant particle number, that is defined as 
\begin{eqnarray}
C_N = T \frac{\partial S}{\partial T}\biggr \vert_N .
\end{eqnarray}
Since we are working in the grand canonical formalism the independent variables are $\mu$ and $T$.  The entropy at constant particle number is found from the chain rule
\begin{eqnarray}
\frac{\partial S}{\partial T}\biggr \vert_N = \frac{\partial S}{\partial \mu}\biggr \vert_T   \frac{\partial \mu}{\partial T}\biggr \vert_N+ \frac{\partial S}{\partial T} \biggr \vert_\mu
\end{eqnarray}
and hence
\begin{eqnarray}
\label{eq:CNtoCRelation}
C_N = T \frac{\partial S}{\partial \mu}\biggr \vert_T   \frac{\partial \mu}{\partial T}\biggr \vert_N+ C.
\end{eqnarray}
It is straightforward to verify that
\begin{eqnarray}
\label{eq:dsdmu}
\frac{\partial S}{\partial \mu}\biggr \vert_T = \beta^2 \left[ \left \langle (E-\mu N) N \right \rangle - \left \langle E-\mu N  \right \rangle \langle N \rangle \right]
\end{eqnarray}
holds. Being a correlator, the right-hand side can be computed within the Monte Carlo simulation. 

In order to obtain ${\partial \mu}/{\partial T}\bigr \vert_N$ it is useful to differentiate $N=\sqrt{T/g'}\psi'(\mu/\sqrt{Tg'})/\psi(\mu/\sqrt{Tg'})$ with respect to $T$, holding $N$ constant. From this one obtains the relation
\begin{eqnarray}
\frac{\partial \mu}{\partial T}\biggr \vert_N  =-\frac{\mu^{2}}{8T^2}\frac{N}{C-\frac{N_{v}+1}{2}+\frac{\mu N}{4T}}+\frac{\mu}{2T}.
\end{eqnarray}

Together with eq. \eqref{eq:dsdmu} this allows the specific heat at constant particle number to be determined from \eqref{eq:CNtoCRelation}. The result is shown in the inset of Fig. \ref{fig:muvsT}.

Differentiating eq. \eqref{eq:genFormS} and substituting the result into eq. \eqref{eq:CNtoCRelation} one can also show that the formula 
\begin{eqnarray}
\label{eq:specHeatNFormula}
C_N = \frac{N_v + 1}{2} + \frac{C - \frac{N_v  + 1}{2}}{C - \frac{N_v + 1}{2} + \frac{\mu N}{4  T}}\frac{\mu N}{4T}
\end{eqnarray}
holds, which provides an alternative way to find $C_N$. 

Using the asymptotics for $C$ from the main text we can find the limiting behavior of $C_N$. The asymptotics for $C_N$ are
\begin{eqnarray}
\label{eq:CNAsymptotics}
\langle C_N \rangle =\begin{cases}
\frac{N_v + 1 + p}{2} & T \rightarrow 0 \\
\frac{N_v + 1}{2} - \frac{b^2}{8 C''(0)}& T \rightarrow T_\star  \\
0 & T  \rightarrow \infty,   \end{cases}
\end{eqnarray} 
where $C''(0) = \left(\psi''(0)/\psi(0) - \left[\psi'(0)/\psi(0)\right]^2\right)/2$. To leading order in the series expansion of $\log \psi(x)$ around zero, $C''(0)$ vanishes, but higher order corrections will generically yield a finite value.
The case where $\mu$ tends to zero, i.e. the limit where $T \rightarrow T_\star$, is peculiar since the denominator vanishes quadratically with $\mu$ and not linearly as one might expect. As a consequence the second term in eq. \eqref{eq:specHeatNFormula} makes a finite contribution as $\mu$ vanishes. 
%%%%%%%%%%%%%%%%%%%%%%%%%%%%%%%%%

\section{Exact partition function for $N_v = 0$}
\label{app:Exact}

\begin{figure}[h]
\centering{}\includegraphics[width=\columnwidth]{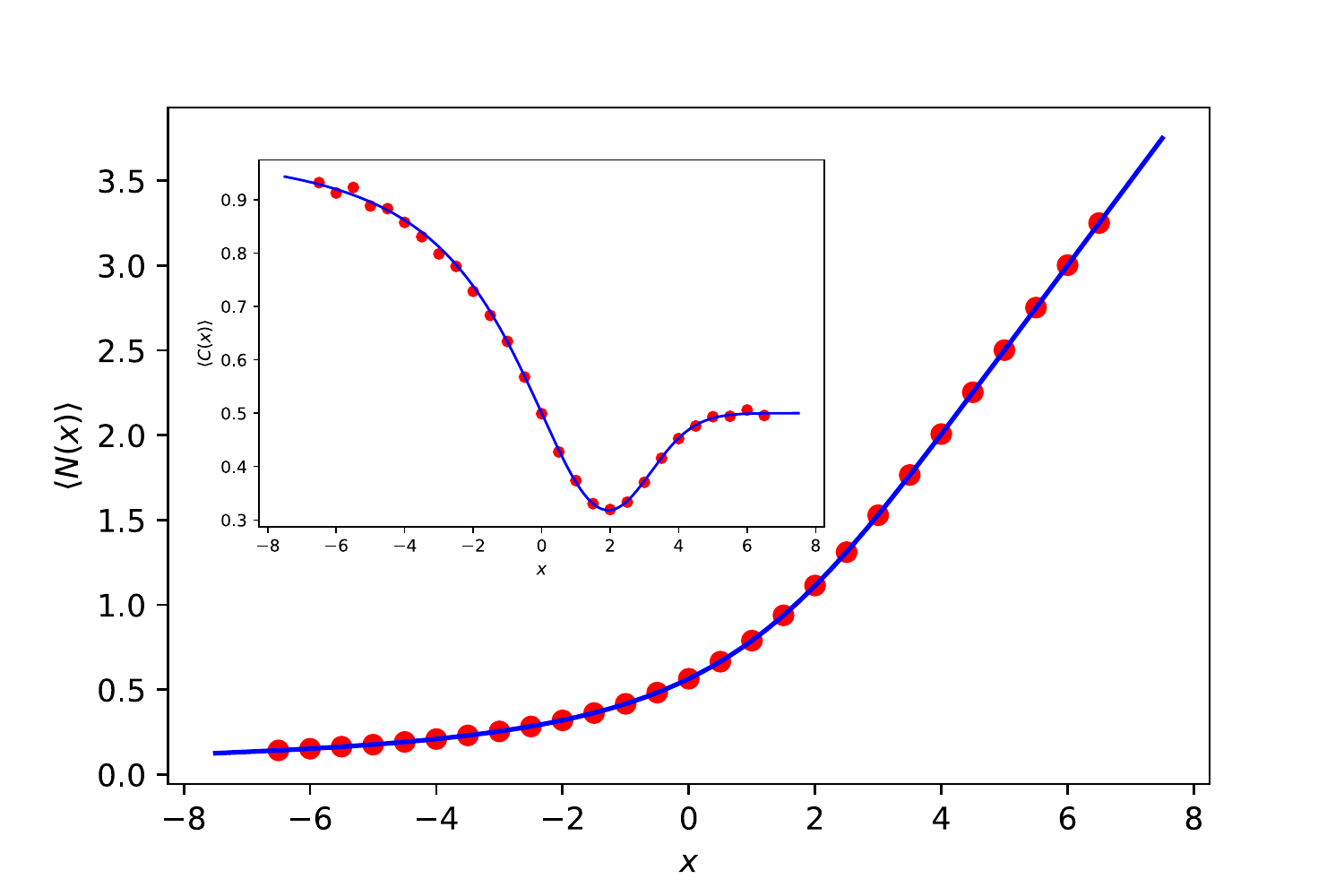}\protect\caption{\label{fig:ExactvsMC} Comparison of Monte Carlo result for the particle number $\langle N \rangle$ on $T = g' , \mu= x g'$ trajectory with the exact result (blue) as given by \eqref{eq:exactN} . The inset shows the MC results for the specific heat together with the exact result (blue) stated in \eqref{eq:exactC}.}
\end{figure}

The case $N_v = 0$ can be solved exactly and thereby provides a check for our Monte-Carlo simulations. Setting ${N_v = 0}$ in equation \eqref{eq:finalHamiltonian} results in a Hamiltonian with only one complex degree of freedom. The complex integral in the partition function \eqref{eq:finalpartitionFct} can be carried out and yields 
\begin{eqnarray}
\label{eq:exactZ}
Z &=& \sqrt{\frac{ T}{g'}} \psi\left(x\right) \nonumber \\
\psi(x)&=&\frac{\sqrt{\pi}}{2} e^{x^2 /4} \left[1+\erf\left(\frac{x}{2}\right) \right],
\end{eqnarray}
with the error-function defined as $\erf(u) = 2/\sqrt{\pi} \int\limits_0^u {e^{-t^2}dt}$. We used the shorthand $x = {\mu}/{\sqrt{g' T}}$, which was introduced in the main text. \newline The particle number $\langle N \rangle$ is given by
\begin{eqnarray}
\label{eq:exactN}
\langle N \rangle=\frac{\mu}{2 g'}+\sqrt{\frac{T}{g'}} \frac{e^{-x^2 /4} }{1+\erf\left(\frac{x}{2}\right)}
\end{eqnarray}
This expression was used in the main text to work out the behavior of the chemical potential at a given value of $\langle N \rangle$. 
The specific heat $\langle C \rangle$ is given by 
\begin{eqnarray}
\label{eq:exactC}
\langle C \rangle&=&\frac{1}{2}-\frac{1}{4\pi}\frac{x^2 }{[1+\erf\left(\frac{x}{2}\right)]^2}e^{-x^2 /2}  \nonumber \\
&&-\frac{1}{8\sqrt{\pi} }\frac{x^3 +2x}{1+\erf\left(\frac{x}{2}\right)} e^{-x^2 /4}.
\end{eqnarray}
Fig. \ref{fig:ExactvsMC} shows that $\langle N \rangle$ and $\langle C \rangle$ as given by these formulas agree very well with the results from the Monte Carlo simulations.
The entropy $S=\partial (T\log Z)/\partial T$ is given by
\begin{eqnarray}
\label{eq:exactS}
\langle S \rangle&=&\frac{1}{2}+\frac{1}{2}\log \frac{\pi}{4}+\frac{1}{2}\log \frac{T}{g'}+\log\left[1+ \erf\left(\frac{x}{2}\right)\right] \nonumber \\
&&- \frac{x}{2}\frac{e^{-\frac{x^2}{4}}}{\sqrt{\pi} \left(2-\erf(x/2)\right)}.
\end{eqnarray}

%%%%%%%%%%%%%%%%%%%%%%%%%%%%%%%%%
\bibliography{library}

\begin{thebibliography}{33}%
\makeatletter
\providecommand \@ifxundefined [1]{%
 \@ifx{#1\undefined}
}%
\providecommand \@ifnum [1]{%
 \ifnum #1\expandafter \@firstoftwo
 \else \expandafter \@secondoftwo
 \fi
}%
\providecommand \@ifx [1]{%
 \ifx #1\expandafter \@firstoftwo
 \else \expandafter \@secondoftwo
 \fi
}%
\providecommand \natexlab [1]{#1}%
\providecommand \enquote  [1]{``#1''}%
\providecommand \bibnamefont  [1]{#1}%
\providecommand \bibfnamefont [1]{#1}%
\providecommand \citenamefont [1]{#1}%
\providecommand \href@noop [0]{\@secondoftwo}%
\providecommand \href [0]{\begingroup \@sanitize@url \@href}%
\providecommand \@href[1]{\@@startlink{#1}\@@href}%
\providecommand \@@href[1]{\endgroup#1\@@endlink}%
\providecommand \@sanitize@url [0]{\catcode `\\12\catcode `\$12\catcode
  `\&12\catcode `\#12\catcode `\^12\catcode `\_12\catcode `\%12\relax}%
\providecommand \@@startlink[1]{}%
\providecommand \@@endlink[0]{}%
\providecommand \url  [0]{\begingroup\@sanitize@url \@url }%
\providecommand \@url [1]{\endgroup\@href {#1}{\urlprefix }}%
\providecommand \urlprefix  [0]{URL }%
\providecommand \Eprint [0]{\href }%
\providecommand \doibase [0]{http://dx.doi.org/}%
\providecommand \selectlanguage [0]{\@gobble}%
\providecommand \bibinfo  [0]{\@secondoftwo}%
\providecommand \bibfield  [0]{\@secondoftwo}%
\providecommand \translation [1]{[#1]}%
\providecommand \BibitemOpen [0]{}%
\providecommand \bibitemStop [0]{}%
\providecommand \bibitemNoStop [0]{.\EOS\space}%
\providecommand \EOS [0]{\spacefactor3000\relax}%
\providecommand \BibitemShut  [1]{\csname bibitem#1\endcsname}%
\let\auto@bib@innerbib\@empty
%</preamble>
\bibitem [{\citenamefont {Fetter}(2009)}]{Fetter2009}%
  \BibitemOpen
  \bibfield  {author} {\bibinfo {author} {\bibfnamefont {A.~L.}\ \bibnamefont
  {Fetter}},\ }\href {\doibase 10.1103/RevModPhys.81.647} {\bibfield  {journal}
  {\bibinfo  {journal} {Rev. Mod. Phys.}\ }\textbf {\bibinfo {volume} {81}},\
  \bibinfo {pages} {647} (\bibinfo {year} {2009})}\BibitemShut {NoStop}%
\bibitem [{\citenamefont {Svistunov}\ \emph {et~al.}(2015)\citenamefont
  {Svistunov}, \citenamefont {Babaev},\ and\ \citenamefont
  {Prokof'ev}}]{svistunov2015superfluid}%
  \BibitemOpen
  \bibfield  {author} {\bibinfo {author} {\bibfnamefont {B.~V.}\ \bibnamefont
  {Svistunov}}, \bibinfo {author} {\bibfnamefont {E.~S.}\ \bibnamefont
  {Babaev}}, \ and\ \bibinfo {author} {\bibfnamefont {N.~V.}\ \bibnamefont
  {Prokof'ev}},\ }\href@noop {} {\emph {\bibinfo {title} {Superfluid states of
  matter}}}\ (\bibinfo  {publisher} {Crc Press},\ \bibinfo {year}
  {2015})\BibitemShut {NoStop}%
\bibitem [{\citenamefont {Sonin}(2016{\natexlab{a}})}]{sonin2016dynamics}%
  \BibitemOpen
  \bibfield  {author} {\bibinfo {author} {\bibfnamefont {E.~B.}\ \bibnamefont
  {Sonin}},\ }\href@noop {} {\emph {\bibinfo {title} {Dynamics of quantised
  vortices in superfluids}}}\ (\bibinfo  {publisher} {Cambridge University
  Press},\ \bibinfo {year} {2016})\BibitemShut {NoStop}%
\bibitem [{\citenamefont {Cooper}(2008)}]{Cooper2008}%
  \BibitemOpen
  \bibfield  {author} {\bibinfo {author} {\bibfnamefont {N.}~\bibnamefont
  {Cooper}},\ }\href {\doibase 10.1080/00018730802564122} {\bibfield  {journal}
  {\bibinfo  {journal} {Advances in Physics}\ }\textbf {\bibinfo {volume}
  {57}},\ \bibinfo {pages} {539} (\bibinfo {year} {2008})},\ \Eprint
  {http://arxiv.org/abs/https://doi.org/10.1080/00018730802564122}
  {https://doi.org/10.1080/00018730802564122} \BibitemShut {NoStop}%
\bibitem [{\citenamefont {Sonin}(2016{\natexlab{b}})}]{Sonin2016}%
  \BibitemOpen
  \bibfield  {author} {\bibinfo {author} {\bibfnamefont {E.~B.}\ \bibnamefont
  {Sonin}},\ }\href {\doibase 10.1017/CBO9781139047616} {\emph {\bibinfo
  {title} {Dynamics of Quantised Vortices in Superfluids}}}\ (\bibinfo
  {publisher} {Cambridge University Press},\ \bibinfo {year}
  {2016})\BibitemShut {NoStop}%
\bibitem [{\citenamefont {Cooper}\ \emph {et~al.}(2001)\citenamefont {Cooper},
  \citenamefont {Wilkin},\ and\ \citenamefont {Gunn}}]{Cooper2001}%
  \BibitemOpen
  \bibfield  {author} {\bibinfo {author} {\bibfnamefont {N.~R.}\ \bibnamefont
  {Cooper}}, \bibinfo {author} {\bibfnamefont {N.~K.}\ \bibnamefont {Wilkin}},
  \ and\ \bibinfo {author} {\bibfnamefont {J.~M.~F.}\ \bibnamefont {Gunn}},\
  }\href {\doibase 10.1103/PhysRevLett.87.120405} {\bibfield  {journal}
  {\bibinfo  {journal} {Phys. Rev. Lett.}\ }\textbf {\bibinfo {volume} {87}},\
  \bibinfo {pages} {120405} (\bibinfo {year} {2001})}\BibitemShut {NoStop}%
\bibitem [{\citenamefont {Sinova}\ \emph {et~al.}(2002)\citenamefont {Sinova},
  \citenamefont {Hanna},\ and\ \citenamefont {MacDonald}}]{Sinova2002}%
  \BibitemOpen
  \bibfield  {author} {\bibinfo {author} {\bibfnamefont {J.}~\bibnamefont
  {Sinova}}, \bibinfo {author} {\bibfnamefont {C.}~\bibnamefont {Hanna}}, \
  and\ \bibinfo {author} {\bibfnamefont {A.~H.}\ \bibnamefont {MacDonald}},\
  }\href {\doibase 10.1103/PhysRevLett.89.030403} {\bibfield  {journal}
  {\bibinfo  {journal} {Phys. Rev. Lett.}\ }\textbf {\bibinfo {volume} {89}},\
  \bibinfo {pages} {030403} (\bibinfo {year} {2002})},\ \Eprint
  {http://arxiv.org/abs/0201020} {0201020 [cond-mat]} \BibitemShut {NoStop}%
\bibitem [{\citenamefont {Viefers}(2008)}]{Viefers2008}%
  \BibitemOpen
  \bibfield  {author} {\bibinfo {author} {\bibfnamefont {S.}~\bibnamefont
  {Viefers}},\ }\href {http://stacks.iop.org/0953-8984/20/i=12/a=123202}
  {\bibfield  {journal} {\bibinfo  {journal} {Journal of Physics: Condensed
  Matter}\ }\textbf {\bibinfo {volume} {20}},\ \bibinfo {pages} {123202}
  (\bibinfo {year} {2008})}\BibitemShut {NoStop}%
\bibitem [{\citenamefont {Abrikosov}(1957)}]{Abrikosov1957}%
  \BibitemOpen
  \bibfield  {author} {\bibinfo {author} {\bibfnamefont {A.~A.}\ \bibnamefont
  {Abrikosov}},\ }\href@noop {} {\bibfield  {journal} {\bibinfo  {journal}
  {Sov. Phys. JETP}\ }\textbf {\bibinfo {volume} {5}},\ \bibinfo {pages} {1174}
  (\bibinfo {year} {1957})}\BibitemShut {NoStop}%
\bibitem [{\citenamefont {Br\'ezin}\ \emph {et~al.}(1985)\citenamefont
  {Br\'ezin}, \citenamefont {Nelson},\ and\ \citenamefont
  {Thiaville}}]{Brezin1985}%
  \BibitemOpen
  \bibfield  {author} {\bibinfo {author} {\bibfnamefont {E.}~\bibnamefont
  {Br\'ezin}}, \bibinfo {author} {\bibfnamefont {D.~R.}\ \bibnamefont
  {Nelson}}, \ and\ \bibinfo {author} {\bibfnamefont {A.}~\bibnamefont
  {Thiaville}},\ }\href {\doibase 10.1103/PhysRevB.31.7124} {\bibfield
  {journal} {\bibinfo  {journal} {Phys. Rev. B}\ }\textbf {\bibinfo {volume}
  {31}},\ \bibinfo {pages} {7124} (\bibinfo {year} {1985})}\BibitemShut
  {NoStop}%
\bibitem [{\citenamefont {Fisher}(1980)}]{Fisher1980}%
  \BibitemOpen
  \bibfield  {author} {\bibinfo {author} {\bibfnamefont {D.~S.}\ \bibnamefont
  {Fisher}},\ }\href {\doibase 10.1103/PhysRevB.22.1190} {\bibfield  {journal}
  {\bibinfo  {journal} {Phys. Rev. B}\ }\textbf {\bibinfo {volume} {22}},\
  \bibinfo {pages} {1190} (\bibinfo {year} {1980})}\BibitemShut {NoStop}%
\bibitem [{\citenamefont {Hikami}\ \emph {et~al.}(1991)\citenamefont {Hikami},
  \citenamefont {Fujita},\ and\ \citenamefont {Larkin}}]{Hikami1991}%
  \BibitemOpen
  \bibfield  {author} {\bibinfo {author} {\bibfnamefont {S.}~\bibnamefont
  {Hikami}}, \bibinfo {author} {\bibfnamefont {A.}~\bibnamefont {Fujita}}, \
  and\ \bibinfo {author} {\bibfnamefont {A.~I.}\ \bibnamefont {Larkin}},\
  }\href {\doibase 10.1103/PhysRevB.44.10400} {\bibfield  {journal} {\bibinfo
  {journal} {Phys. Rev. B}\ }\textbf {\bibinfo {volume} {44}},\ \bibinfo
  {pages} {10400} (\bibinfo {year} {1991})}\BibitemShut {NoStop}%
\bibitem [{\citenamefont {Te\ifmmode \check{s}\else
  \v{s}\fi{}anovi\ifmmode~\acute{c}\else \'{c}\fi{}}\ \emph
  {et~al.}(1992)\citenamefont {Te\ifmmode \check{s}\else
  \v{s}\fi{}anovi\ifmmode~\acute{c}\else \'{c}\fi{}}, \citenamefont {Xing},
  \citenamefont {Bulaevskii}, \citenamefont {Li},\ and\ \citenamefont
  {Suenaga}}]{Tesanovic1992}%
  \BibitemOpen
  \bibfield  {author} {\bibinfo {author} {\bibfnamefont {Z.}~\bibnamefont
  {Te\ifmmode \check{s}\else \v{s}\fi{}anovi\ifmmode~\acute{c}\else
  \'{c}\fi{}}}, \bibinfo {author} {\bibfnamefont {L.}~\bibnamefont {Xing}},
  \bibinfo {author} {\bibfnamefont {L.}~\bibnamefont {Bulaevskii}}, \bibinfo
  {author} {\bibfnamefont {Q.}~\bibnamefont {Li}}, \ and\ \bibinfo {author}
  {\bibfnamefont {M.}~\bibnamefont {Suenaga}},\ }\href {\doibase
  10.1103/PhysRevLett.69.3563} {\bibfield  {journal} {\bibinfo  {journal}
  {Phys. Rev. Lett.}\ }\textbf {\bibinfo {volume} {69}},\ \bibinfo {pages}
  {3563} (\bibinfo {year} {1992})}\BibitemShut {NoStop}%
\bibitem [{\citenamefont {Kato}\ and\ \citenamefont
  {Nagaosa}(1993)}]{Kato1993}%
  \BibitemOpen
  \bibfield  {author} {\bibinfo {author} {\bibfnamefont {Y.}~\bibnamefont
  {Kato}}\ and\ \bibinfo {author} {\bibfnamefont {N.}~\bibnamefont {Nagaosa}},\
  }\href {\doibase 10.1103/PhysRevB.48.7383} {\bibfield  {journal} {\bibinfo
  {journal} {Phys. Rev. B}\ }\textbf {\bibinfo {volume} {48}},\ \bibinfo
  {pages} {7383} (\bibinfo {year} {1993})}\BibitemShut {NoStop}%
\bibitem [{\citenamefont {O'Neill}\ and\ \citenamefont
  {Moore}(1993)}]{ONeil1993}%
  \BibitemOpen
  \bibfield  {author} {\bibinfo {author} {\bibfnamefont {J.~A.}\ \bibnamefont
  {O'Neill}}\ and\ \bibinfo {author} {\bibfnamefont {M.~A.}\ \bibnamefont
  {Moore}},\ }\href {\doibase 10.1103/PhysRevB.48.374} {\bibfield  {journal}
  {\bibinfo  {journal} {Phys. Rev. B}\ }\textbf {\bibinfo {volume} {48}},\
  \bibinfo {pages} {374} (\bibinfo {year} {1993})}\BibitemShut {NoStop}%
\bibitem [{\citenamefont {Roy}\ \emph {et~al.}(2019)\citenamefont {Roy},
  \citenamefont {Dutta}, \citenamefont {Roy~Choudhury}, \citenamefont
  {Basistha}, \citenamefont {Maccari}, \citenamefont {Mandal}, \citenamefont
  {Jesudasan}, \citenamefont {Bagwe}, \citenamefont {Castellani}, \citenamefont
  {Benfatto},\ and\ \citenamefont {Raychaudhuri}}]{Roy2019}%
  \BibitemOpen
  \bibfield  {author} {\bibinfo {author} {\bibfnamefont {I.}~\bibnamefont
  {Roy}}, \bibinfo {author} {\bibfnamefont {S.}~\bibnamefont {Dutta}}, \bibinfo
  {author} {\bibfnamefont {A.~N.}\ \bibnamefont {Roy~Choudhury}}, \bibinfo
  {author} {\bibfnamefont {S.}~\bibnamefont {Basistha}}, \bibinfo {author}
  {\bibfnamefont {I.}~\bibnamefont {Maccari}}, \bibinfo {author} {\bibfnamefont
  {S.}~\bibnamefont {Mandal}}, \bibinfo {author} {\bibfnamefont
  {J.}~\bibnamefont {Jesudasan}}, \bibinfo {author} {\bibfnamefont
  {V.}~\bibnamefont {Bagwe}}, \bibinfo {author} {\bibfnamefont
  {C.}~\bibnamefont {Castellani}}, \bibinfo {author} {\bibfnamefont
  {L.}~\bibnamefont {Benfatto}}, \ and\ \bibinfo {author} {\bibfnamefont
  {P.}~\bibnamefont {Raychaudhuri}},\ }\href {\doibase
  10.1103/PhysRevLett.122.047001} {\bibfield  {journal} {\bibinfo  {journal}
  {Phys. Rev. Lett.}\ }\textbf {\bibinfo {volume} {122}},\ \bibinfo {pages}
  {047001} (\bibinfo {year} {2019})}\BibitemShut {NoStop}%
\bibitem [{\citenamefont {Gifford}\ and\ \citenamefont
  {Baym}(2008)}]{Gifford2008}%
  \BibitemOpen
  \bibfield  {author} {\bibinfo {author} {\bibfnamefont {S.~A.}\ \bibnamefont
  {Gifford}}\ and\ \bibinfo {author} {\bibfnamefont {G.}~\bibnamefont {Baym}},\
  }\href {\doibase 10.1103/PhysRevA.78.043607} {\bibfield  {journal} {\bibinfo
  {journal} {Phys. Rev. A}\ }\textbf {\bibinfo {volume} {78}},\ \bibinfo
  {pages} {043607} (\bibinfo {year} {2008})}\BibitemShut {NoStop}%
\bibitem [{\citenamefont {Bourne}\ \emph {et~al.}(2007)\citenamefont {Bourne},
  \citenamefont {Gunn},\ and\ \citenamefont {Wilkin}}]{PhysRevA.76.053602}%
  \BibitemOpen
  \bibfield  {author} {\bibinfo {author} {\bibfnamefont {A.}~\bibnamefont
  {Bourne}}, \bibinfo {author} {\bibfnamefont {J.~M.~F.}\ \bibnamefont {Gunn}},
  \ and\ \bibinfo {author} {\bibfnamefont {N.~K.}\ \bibnamefont {Wilkin}},\
  }\href {\doibase 10.1103/PhysRevA.76.053602} {\bibfield  {journal} {\bibinfo
  {journal} {Phys. Rev. A}\ }\textbf {\bibinfo {volume} {76}},\ \bibinfo
  {pages} {053602} (\bibinfo {year} {2007})}\BibitemShut {NoStop}%
\bibitem [{\citenamefont {Saarikoski}\ \emph {et~al.}(2010)\citenamefont
  {Saarikoski}, \citenamefont {Reimann}, \citenamefont {Harju},\ and\
  \citenamefont {Manninen}}]{saarikoski2010vortices}%
  \BibitemOpen
  \bibfield  {author} {\bibinfo {author} {\bibfnamefont {H.}~\bibnamefont
  {Saarikoski}}, \bibinfo {author} {\bibfnamefont {S.}~\bibnamefont {Reimann}},
  \bibinfo {author} {\bibfnamefont {A.}~\bibnamefont {Harju}}, \ and\ \bibinfo
  {author} {\bibfnamefont {M.}~\bibnamefont {Manninen}},\ }\href@noop {}
  {\bibfield  {journal} {\bibinfo  {journal} {Reviews of Modern Physics}\
  }\textbf {\bibinfo {volume} {82}},\ \bibinfo {pages} {2785} (\bibinfo {year}
  {2010})}\BibitemShut {NoStop}%
\bibitem [{\citenamefont {Nascimb{\`e}ne}\ \emph {et~al.}(2010)\citenamefont
  {Nascimb{\`e}ne}, \citenamefont {Navon}, \citenamefont {Jiang}, \citenamefont
  {Chevy},\ and\ \citenamefont {Salomon}}]{Nascimbene2010}%
  \BibitemOpen
  \bibfield  {author} {\bibinfo {author} {\bibfnamefont {S.}~\bibnamefont
  {Nascimb{\`e}ne}}, \bibinfo {author} {\bibfnamefont {N.}~\bibnamefont
  {Navon}}, \bibinfo {author} {\bibfnamefont {K.}~\bibnamefont {Jiang}},
  \bibinfo {author} {\bibfnamefont {F.}~\bibnamefont {Chevy}}, \ and\ \bibinfo
  {author} {\bibfnamefont {C.}~\bibnamefont {Salomon}},\ }\href@noop {}
  {\bibfield  {journal} {\bibinfo  {journal} {Nature}\ }\textbf {\bibinfo
  {volume} {463}},\ \bibinfo {pages} {1057} (\bibinfo {year}
  {2010})}\BibitemShut {NoStop}%
\bibitem [{\citenamefont {Yefsah}\ \emph {et~al.}(2011)\citenamefont {Yefsah},
  \citenamefont {Desbuquois}, \citenamefont {Chomaz}, \citenamefont
  {G\"unter},\ and\ \citenamefont {Dalibard}}]{Yefsah2011}%
  \BibitemOpen
  \bibfield  {author} {\bibinfo {author} {\bibfnamefont {T.}~\bibnamefont
  {Yefsah}}, \bibinfo {author} {\bibfnamefont {R.}~\bibnamefont {Desbuquois}},
  \bibinfo {author} {\bibfnamefont {L.}~\bibnamefont {Chomaz}}, \bibinfo
  {author} {\bibfnamefont {K.~J.}\ \bibnamefont {G\"unter}}, \ and\ \bibinfo
  {author} {\bibfnamefont {J.}~\bibnamefont {Dalibard}},\ }\href {\doibase
  10.1103/PhysRevLett.107.130401} {\bibfield  {journal} {\bibinfo  {journal}
  {Phys. Rev. Lett.}\ }\textbf {\bibinfo {volume} {107}},\ \bibinfo {pages}
  {130401} (\bibinfo {year} {2011})}\BibitemShut {NoStop}%
\bibitem [{\citenamefont {Ku}\ \emph {et~al.}(2012)\citenamefont {Ku},
  \citenamefont {Sommer}, \citenamefont {Cheuk},\ and\ \citenamefont
  {Zwierlein}}]{Ku2012}%
  \BibitemOpen
  \bibfield  {author} {\bibinfo {author} {\bibfnamefont {M.~J.~H.}\
  \bibnamefont {Ku}}, \bibinfo {author} {\bibfnamefont {A.~T.}\ \bibnamefont
  {Sommer}}, \bibinfo {author} {\bibfnamefont {L.~W.}\ \bibnamefont {Cheuk}}, \
  and\ \bibinfo {author} {\bibfnamefont {M.~W.}\ \bibnamefont {Zwierlein}},\
  }\href {\doibase 10.1126/science.1214987} {\bibfield  {journal} {\bibinfo
  {journal} {Science}\ }\textbf {\bibinfo {volume} {335}},\ \bibinfo {pages}
  {563} (\bibinfo {year} {2012})},\ \Eprint
  {http://arxiv.org/abs/https://science.sciencemag.org/content/335/6068/563.full.pdf}
  {https://science.sciencemag.org/content/335/6068/563.full.pdf} \BibitemShut
  {NoStop}%
\bibitem [{\citenamefont {Desbuquois}\ \emph {et~al.}(2014)\citenamefont
  {Desbuquois}, \citenamefont {Yefsah}, \citenamefont {Chomaz}, \citenamefont
  {Weitenberg}, \citenamefont {Corman}, \citenamefont {Nascimb\`ene},\ and\
  \citenamefont {Dalibard}}]{Desbuquois2014}%
  \BibitemOpen
  \bibfield  {author} {\bibinfo {author} {\bibfnamefont {R.}~\bibnamefont
  {Desbuquois}}, \bibinfo {author} {\bibfnamefont {T.}~\bibnamefont {Yefsah}},
  \bibinfo {author} {\bibfnamefont {L.}~\bibnamefont {Chomaz}}, \bibinfo
  {author} {\bibfnamefont {C.}~\bibnamefont {Weitenberg}}, \bibinfo {author}
  {\bibfnamefont {L.}~\bibnamefont {Corman}}, \bibinfo {author} {\bibfnamefont
  {S.}~\bibnamefont {Nascimb\`ene}}, \ and\ \bibinfo {author} {\bibfnamefont
  {J.}~\bibnamefont {Dalibard}},\ }\href {\doibase
  10.1103/PhysRevLett.113.020404} {\bibfield  {journal} {\bibinfo  {journal}
  {Phys. Rev. Lett.}\ }\textbf {\bibinfo {volume} {113}},\ \bibinfo {pages}
  {020404} (\bibinfo {year} {2014})}\BibitemShut {NoStop}%
\bibitem [{\citenamefont {Bloch}\ \emph {et~al.}(2008)\citenamefont {Bloch},
  \citenamefont {Dalibard},\ and\ \citenamefont {Zwerger}}]{Bloch2008}%
  \BibitemOpen
  \bibfield  {author} {\bibinfo {author} {\bibfnamefont {I.}~\bibnamefont
  {Bloch}}, \bibinfo {author} {\bibfnamefont {J.}~\bibnamefont {Dalibard}}, \
  and\ \bibinfo {author} {\bibfnamefont {W.}~\bibnamefont {Zwerger}},\ }\href
  {\doibase 10.1103/RevModPhys.80.885} {\bibfield  {journal} {\bibinfo
  {journal} {Rev. Mod. Phys.}\ }\textbf {\bibinfo {volume} {80}},\ \bibinfo
  {pages} {885} (\bibinfo {year} {2008})}\BibitemShut {NoStop}%
\bibitem [{\citenamefont {Aidelsburger}\ \emph {et~al.}(2018)\citenamefont
  {Aidelsburger}, \citenamefont {Nascimbene},\ and\ \citenamefont
  {Goldman}}]{Aidelsburger2018}%
  \BibitemOpen
  \bibfield  {author} {\bibinfo {author} {\bibfnamefont {M.}~\bibnamefont
  {Aidelsburger}}, \bibinfo {author} {\bibfnamefont {S.}~\bibnamefont
  {Nascimbene}}, \ and\ \bibinfo {author} {\bibfnamefont {N.}~\bibnamefont
  {Goldman}},\ }\href {\doibase https://doi.org/10.1016/j.crhy.2018.03.002}
  {\bibfield  {journal} {\bibinfo  {journal} {Comptes Rendus Physique}\
  }\textbf {\bibinfo {volume} {19}},\ \bibinfo {pages} {394 } (\bibinfo {year}
  {2018})}\BibitemShut {NoStop}%
\bibitem [{\citenamefont {Schweikhard}\ \emph {et~al.}(2004)\citenamefont
  {Schweikhard}, \citenamefont {Coddington}, \citenamefont {Engels},
  \citenamefont {Mogendorff},\ and\ \citenamefont
  {Cornell}}]{schweikhard2004rapidly}%
  \BibitemOpen
  \bibfield  {author} {\bibinfo {author} {\bibfnamefont {V.}~\bibnamefont
  {Schweikhard}}, \bibinfo {author} {\bibfnamefont {I.}~\bibnamefont
  {Coddington}}, \bibinfo {author} {\bibfnamefont {P.}~\bibnamefont {Engels}},
  \bibinfo {author} {\bibfnamefont {V.}~\bibnamefont {Mogendorff}}, \ and\
  \bibinfo {author} {\bibfnamefont {E.~A.}\ \bibnamefont {Cornell}},\
  }\href@noop {} {\bibfield  {journal} {\bibinfo  {journal} {Physical review
  letters}\ }\textbf {\bibinfo {volume} {92}},\ \bibinfo {pages} {040404}
  (\bibinfo {year} {2004})}\BibitemShut {NoStop}%
\bibitem [{\citenamefont {Abo-Shaeer}\ \emph {et~al.}(2001)\citenamefont
  {Abo-Shaeer}, \citenamefont {Raman}, \citenamefont {Vogels},\ and\
  \citenamefont {Ketterle}}]{Abo2001}%
  \BibitemOpen
  \bibfield  {author} {\bibinfo {author} {\bibfnamefont {J.}~\bibnamefont
  {Abo-Shaeer}}, \bibinfo {author} {\bibfnamefont {C.}~\bibnamefont {Raman}},
  \bibinfo {author} {\bibfnamefont {J.}~\bibnamefont {Vogels}}, \ and\ \bibinfo
  {author} {\bibfnamefont {W.}~\bibnamefont {Ketterle}},\ }\href@noop {}
  {\bibfield  {journal} {\bibinfo  {journal} {Science}\ }\textbf {\bibinfo
  {volume} {292}},\ \bibinfo {pages} {476} (\bibinfo {year}
  {2001})}\BibitemShut {NoStop}%
\bibitem [{\citenamefont {Fletcher}\ \emph {et~al.}(2019)\citenamefont
  {Fletcher}, \citenamefont {Shaffer}, \citenamefont {Wilson}, \citenamefont
  {Patel}, \citenamefont {Yan}, \citenamefont {Crépel}, \citenamefont
  {Mukherjee},\ and\ \citenamefont {Zwierlein}}]{Fletcher2019}%
  \BibitemOpen
  \bibfield  {author} {\bibinfo {author} {\bibfnamefont {R.}~\bibnamefont
  {Fletcher}}, \bibinfo {author} {\bibfnamefont {A.}~\bibnamefont {Shaffer}},
  \bibinfo {author} {\bibfnamefont {C.}~\bibnamefont {Wilson}}, \bibinfo
  {author} {\bibfnamefont {P.}~\bibnamefont {Patel}}, \bibinfo {author}
  {\bibfnamefont {Z.}~\bibnamefont {Yan}}, \bibinfo {author} {\bibfnamefont
  {V.}~\bibnamefont {Crépel}}, \bibinfo {author} {\bibfnamefont
  {B.}~\bibnamefont {Mukherjee}}, \ and\ \bibinfo {author} {\bibfnamefont
  {M.}~\bibnamefont {Zwierlein}},\ }\href@noop {} {\bibfield  {journal}
  {\bibinfo  {journal} {arXiv:1911.12347}\ } (\bibinfo {year}
  {2019})}\BibitemShut {NoStop}%
\bibitem [{\citenamefont {Chalopin}\ \emph {et~al.}(2020)\citenamefont
  {Chalopin}, \citenamefont {Satoor}, \citenamefont {Evrard}, \citenamefont
  {Makhalov}, \citenamefont {Dalibard}, \citenamefont {Lopes},\ and\
  \citenamefont {Nascimbene}}]{Chalopin2020}%
  \BibitemOpen
  \bibfield  {author} {\bibinfo {author} {\bibfnamefont {T.}~\bibnamefont
  {Chalopin}}, \bibinfo {author} {\bibfnamefont {T.}~\bibnamefont {Satoor}},
  \bibinfo {author} {\bibfnamefont {A.}~\bibnamefont {Evrard}}, \bibinfo
  {author} {\bibfnamefont {V.}~\bibnamefont {Makhalov}}, \bibinfo {author}
  {\bibfnamefont {J.}~\bibnamefont {Dalibard}}, \bibinfo {author}
  {\bibfnamefont {R.}~\bibnamefont {Lopes}}, \ and\ \bibinfo {author}
  {\bibfnamefont {S.}~\bibnamefont {Nascimbene}},\ }\href@noop {} {\bibfield
  {journal} {\bibinfo  {journal} {arXiv:2001.01664}\ } (\bibinfo {year}
  {2020})}\BibitemShut {NoStop}%
\bibitem [{Note1()}]{Note1}%
  \BibitemOpen
  \bibinfo {note} {At the cost of a Vandermonde determinant one could describe
  the partition function in terms of vortex coordinates instead of the
  coefficients $\protect \{c_n\protect \}$. However, in this case the contact
  interaction between bosons leads to a rather complicated
  multivortex-interaction \cite {PhysRevA.76.053602}. For this reason we prefer
  to work within the $\protect \{c_n\protect \}$-description, and only make
  reference to vortices when discussing the vortex crystal at low
  temperatures.}\BibitemShut {Stop}%
\bibitem [{sup()}]{supmat}%
  \BibitemOpen
  \href@noop {} {}\bibinfo {howpublished} {Supplemental Material contains a
  video showing the melting and crystallization of the vortex lattice
  as seen in Monte Carlo snapshots. It can be accessed via this \href{https://youtu.be/4-nOKptY75w}{YouTube link}. }\BibitemShut {Stop}%
\bibitem [{\citenamefont {Fermi}(1956)}]{FermiThermo}%
  \BibitemOpen
  \bibfield  {author} {\bibinfo {author} {\bibfnamefont {E.}~\bibnamefont
  {Fermi}},\ }\href@noop {} {\emph {\bibinfo {title} {Thermodynamics}}}\
  (\bibinfo  {publisher} {Dover},\ \bibinfo {year} {1956})\BibitemShut
  {NoStop}%
\bibitem [{\citenamefont {Bernard}\ and\ \citenamefont
  {Krauth}(2011)}]{PhysRevLett.107.155704}%
  \BibitemOpen
  \bibfield  {author} {\bibinfo {author} {\bibfnamefont {E.~P.}\ \bibnamefont
  {Bernard}}\ and\ \bibinfo {author} {\bibfnamefont {W.}~\bibnamefont
  {Krauth}},\ }\href {\doibase 10.1103/PhysRevLett.107.155704} {\bibfield
  {journal} {\bibinfo  {journal} {Phys. Rev. Lett.}\ }\textbf {\bibinfo
  {volume} {107}},\ \bibinfo {pages} {155704} (\bibinfo {year}
  {2011})}\BibitemShut {NoStop}%
\end{thebibliography}%

\end{document}